\documentclass[%
 reprint,
 superscriptaddress,
 nofootinbib,
 twocolumn, 
 amsmath,amssymb,
 aps,
 prl,
]{revtex4-2}

\usepackage[colorlinks = true,
            linkcolor = blue,
            urlcolor  = red,
            citecolor = red,
            anchorcolor = blue]{hyperref}
\usepackage[T1]{fontenc}
\usepackage[utf8]{inputenc}
\usepackage{graphicx}
\usepackage{dcolumn}
\usepackage{color}
\usepackage{amsmath}
\usepackage{braket}
\usepackage{amssymb}
\usepackage{appendix}
\usepackage{MnSymbol}
\usepackage{todonotes}
\usepackage{orcidlink}

\usepackage{glossaries}
\newacronym{CDW}{CDW}{charge-density-wave}
\newacronym{SC}{SC}{spin-conserving}
\newacronym{NSC}{NSC}{non-spin-conserving}
\newacronym{RIXS}{RIXS}{resonant inelastic x-ray scattering}
\newacronym{DMRG}{DMRG}{density matrix renormalization group}
\newacronym{BOW}{BOW}{bond-order-wave}
\newacronym{DQMC}{DQMC}{determinant quantum Monte Carlo}
\newacronym{QMC}{QMC}{quantum Monte Carlo}
\newacronym{SSH}{SSH}{Su-Schrieffer-Heeger}
\newacronym{2D}{2D}{two-dimension}
\newacronym{1D}{1D}{one-dimensional}
\newacronym{FS}{FS}{Fermi surface}
\newacronym{ED}{ED}{exact diagonalization}
\newacronym{eph}{$e$-ph}{electron-phonon}
\newacronym{HSSH}{HSSH}{Hubbard-SSH}
\newacronym{XAS}{XAS}{x-ray absorption spectroscopy}
\newacronym{ee}{$e$-$e$}{electron-electron}
\newacronym{AFM}{AFM}{antiferromagnetism}
\newacronym{PSF}{PSF}{phonon spectral function}
\newacronym{SOM}{SOM}{supplementary online materials}
\newacronym{INS}{INS}{inelastic neutron scattering}
\newacronym{HH}{HH}{Hubbard-Holstein}

\begin{document}

\preprint{}
\title{Identifying and quantifying {S}u-{S}chrieffer-{H}eeger-like interactions with {RIXS}}

\author{Debshikha Banerjee\,\orcidlink{0009-0001-2925-9724}}
\affiliation{Department of Physics and Astronomy, The University of Tennessee, Knoxville, Tennessee 37996, USA}
\affiliation{Institute for Advanced Materials and Manufacturing, University of Tennessee, Knoxville, Tennessee 37996, USA\looseness=-1}
\author{Jinu Thomas\,\orcidlink{0000-0003-4818-6660}}
\affiliation{Department of Physics and Astronomy, The University of Tennessee, 
Knoxville, Tennessee 37996, USA}
\affiliation{Institute for Advanced Materials and Manufacturing, University of Tennessee, Knoxville, Tennessee 37996, USA\looseness=-1}
\author{Alberto Nocera\,\orcidlink{0000-0001-9722-6388}}
\affiliation{Stewart Blusson Quantum Matter Institute, University of British Columbia, Vancouver, British Columbia, Canada
V6T 1Z4}
\affiliation{Department of Physics Astronomy, University of British Columbia, Vancouver, British Columbia, Canada V6T 1Z1}
\author{Steven Johnston\,\orcidlink{0000-0002-2343-0113}}
\affiliation{Department of Physics and Astronomy, The University of Tennessee, Knoxville, Tennessee 37996, USA}
\affiliation{Institute for Advanced Materials and Manufacturing, University of Tennessee, Knoxville, Tennessee 37996, USA\looseness=-1}

\date{\today}

\begin{abstract}
Su-Schrieffer-Heeger (SSH)-like electron-phonon ($e$-ph) interactions can drive the formation of light (bi)polarons and several novel states of matter. It is, therefore, prudent to develop experimental protocols for identifying such couplings in real materials and quantifying their strength. Here, we investigate how resonant inelastic x-ray scattering (RIXS) probes $e$-ph interactions in the one-dimensional half-filled Hubbard-SSH model with onsite phonons. Using the density matrix renormalization group method, we compute the full RIXS response and find that the lattice excitations generated during the scattering process inevitably couple to the system's charge and magnetic sectors, resulting in combined multi-particle excitations that cannot be easily disentangled from one another. While this aspect complicates the interpretation of RIXS experiments, we outline how it can be leveraged to identify and quantify SSH-like interactions in quantum materials. 
\end{abstract}

\maketitle

\noindent{\bf Introduction}. There is a growing recognition that short-range \gls*{SSH}-like interactions, where the atomic motion modulates the system's hopping integrals~\cite{Barisic1970tightbinding, Su1979solitons}, can produce results counter to conventional wisdom for \gls*{eph} coupling. For example, these interactions have been linked to concepts in topology~\cite{Jackiw1976Solitons, Heeger1988solitons, meier2016observation, bid2022topological} and help stabilize and control the properties of Dirac and Weyl semimetals~\cite{Moeller2017typeII}. 
They can also drive \gls*{CDW}~\cite{li2020quantum, CohenStead2023hybrid}, \gls*{BOW}~\cite{Feng2022phase, Goetz2022valence, MalkarugeCosta2024Kukule}, antiferromagnetic~\cite{Goetz2022valence, Cai2022robustness} 
and quantum spin liquid~\cite{cai2024quantumspinliquidelectronphonon} phases, or produce light bipolarons~\cite{Sous2018light, Nocera2021bipolaron, TanjaroonLy2023comparative} with the potential for high-temperature superconductivity~\cite{Zhang2023bipolaronic, cai2023hightemperature}. Given this rich landscape, a key challenge is to create experimental protocols for identifying and quantifying \gls*{SSH} interactions in materials. 

\Gls*{RIXS} is a powerful probe of quantum materials~\cite{Mitriano2024exploring}, which has experienced significant growth as a tool for studying \gls*{eph} interactions~\cite{Ament2011determining, hancock2010lattice, Yavas2010observation, Lee2013Role, Devereaux2016directly, Rossi2019experimental, Lin2020strongly, Peng2020enhanced, Braicovich2020determining, Dashwood2021probing, Gilmore2023quantifying}. Motivated by this, here we present a detailed \gls*{DMRG} study of the \gls*{RIXS} spectra of a \gls*{1D} \gls*{HSSH} chain with coupling to optical phonons~\cite{Capone1997small}. By computing the full \gls*{RIXS} response on extended chains~\cite{Nocera2018Computing}, we capture the relevant lattice, spin, and charge fluctuations and their important interplay~\cite{Bieniasz2021beyond, Thomas2024theory}. We demonstrate that the phonon excitations are intrinsically coupled to the system's spin and charge sectors, causing them to appear at energies below the dressed phonon energies. These observations differ significantly from the predictions of both the single-site~\cite{Ament2011resonant} and Feynman diagram~\cite{Devereaux2016directly} approaches to the problem and complicate the interpretation of experiments. Nevertheless, they also provide a unique fingerprint of \gls*{SSH}-like interactions that can be used to identify and quantify their presence in a material. While our results are obtained in \gls*{1D}, the principles identified here are general and should apply in higher dimensions.

\noindent{\bf Model \& Methods}.  
We study the \gls*{1D} \gls*{HSSH} Hamiltonian~\cite{Capone1997small, MalkarugeCosta2023comparative}
\begin{align}\label{eq:Hamiltonian} 
    H=&-t \sum_{j,\sigma} \left[
    c^\dagger_{j,\sigma}c^{\phantom\dagger}_{j+1,\sigma} + \text{H.c.}\right] 
    + \hbar\omega_\text{ph} \sum_{j}\left(b^\dagger_j b^{\phantom\dagger}_j+\tfrac{1}{2}\right) \\\nonumber
    &+ U \sum_{j}\hat{n}_{j,\uparrow}\hat{n}_{j,\downarrow} + 
   g\sum_{j,\sigma} \left[c^\dagger_{j,\sigma}c^{\phantom\dagger}_{j+1,\sigma}\left(\hat{X}_{j+1} - \hat{X}_{j}\right) + \text{h.c.}\right].
\end{align}
Here $c^\dagger_{j,\sigma}$ ($c^{\phantom\dagger}_{j,\sigma}$) creates (annihilates) a spin-$\sigma$ electron on lattice site $j$, $b^\dagger_j$ ($b^{\phantom\dagger}_{j}$) creates (annihilates) a phonon mode with energy $\omega_\text{ph}$ at site $j$, $\hat{X}_j = \sqrt{\hbar/2M\omega_\text{ph}}(b^\dagger_j + b^{\phantom\dagger}_j)$ is the displacement operator, $\hat{n}_{j,\sigma} = c^\dagger_{j,\sigma}c^{\phantom\dagger}_{j,\sigma}$ is the electron number operator, $t$ is the nearest-neighbor hopping, $U$ is the Hubbard repulsion, and $g$ is the \gls*{eph} coupling strength. Throughout, we consider $L$-site chains at half-filling with open boundary conditions and set $\hbar = M = t = \omega_\text{ph} = 1$. 

Our model is relevant for systems where only one valence orbital is active in the \gls*{RIXS} process, e.g., transition $L$-edge measurements on $d^9$ systems. The \gls*{RIXS} intensity is given by 
\begin{equation}
    I(q,\Omega) \propto \sum_f \: \lvert \mathcal{F}_{fi} \rvert^2 \delta(E_f -E_i-\Omega) 
    \label{eq:intensity}
\end{equation}
with the scattering amplitude
\begin{equation}
    \mathcal{F}_{fi} = \sum_{{n,j,\sigma^\prime,\sigma}} e^{\mathrm{i} q R_j} \frac{\bra{f}D_{j,\sigma^\prime}^{\dagger}\ket{n}\bra{n}D_{j,\sigma}^{\phantom\dagger}\ket{i}}{E_i +\omega_\mathrm{in}-E_n+\mathrm{i}\Gamma/2}. 
    \label{eq:amplitude}
\end{equation}
Here, $\omega_\mathrm{in}$ ($\omega_\mathrm{out}$) and $k_\mathrm{in}$ ($k_\mathrm{out}$)
are the incoming (outgoing) photon energies and momenta, respectively; $\Omega=\omega_\mathrm{in} - \omega_\mathrm{out}$ $(q = k_\mathrm{in}-k_\mathrm{out})$ is the energy (momentum) transfer; $\ket{i}$, $\ket{n}$, and $\ket{f}$ are the initial, intermediate, and final states of the RIXS process with energies $E_i$, $E_n$, and $E_f$, respectively; $\Gamma/2$ is the inverse core-hole lifetime; and $D_{j,\sigma} = \left(c_{j,\sigma}^{\dagger}p^{\phantom\dagger}_{j,\alpha} + \text{h.c.}\right)$ is the local dipole operator for the atomic transition, where $p_{j,\alpha}$ annihilates a core electron in energy level $\alpha$ at site $j$ located at position $R_j$. (Here, we have neglected the orbital matrix element to be consistent with prior work examining lattice excitations~\cite{Devereaux2016directly, Thomas2024theory}.) 
If there is a strong spin-orbit coupling in the core orbitals, then $\alpha$ corresponds to the total angular momentum, and the electron spin is not a good quantum number~\cite{Braicovich2009dispersion}. We have found that the phonon excitations are strongest in the \gls*{SC} channel~\cite{Supplement} and so we focus on this channel in the main text by restricting $\sigma = \sigma^\prime$ in Eq.~\eqref{eq:amplitude}. Importantly, this channel can be isolated from the 
\gls*{NSC} channel ($\sigma \ne \sigma^\prime$) in \gls*{1D} systems using a polarimetry analysis~\cite{Bisogni2014femtosecond}.

We evaluate Eq.~\eqref{eq:intensity} using the DMRG++ code~\cite{Alvarez2009density} as described in Ref.~\cite{Nocera2018Computing}. We keep up to $m=500$ \gls*{DMRG} states while restricting the local phonon Hilbert space to keep $N_\text{p} = 10-13$ phonon modes per site, depending on the model parameters. We have verified that our results are converged with respect to $m$ and $N_\text{p}$~\cite{Supplement}. While calculating the intermediate state, we include the interaction with the core-hole via an additional term 
$H_\mathrm{c} = V_\mathrm{c}\sum_{j,\sigma}\hat{c}^{\dagger}_{j,\sigma}\hat{c}^{\phantom \dagger}_{j,\sigma}(1-
p^\dagger_{j,\alpha}p^{\phantom\dagger}_{j,\alpha})$, where $V_\mathrm{c}=-8t$ is the core-hole potential, and set $\Gamma/2 = t/4$ and a final state broadening $\eta = 0.2t$. Our conclusions are not sensitive to the precise values of $V_\mathrm{c}$ and $\Gamma/2$~\cite{Supplement}. Finally, we also calculate the \gls*{XAS} spectra, phonon spectral function $B(q,\omega)$, and dynamical spin  $S(q,\omega)$ and charge $N(q,\omega)$ structure factors, as described in Ref.~\cite{Supplement}.  \\

\begin{figure}[t]
    \centering
    \includegraphics[width=0.85\columnwidth]{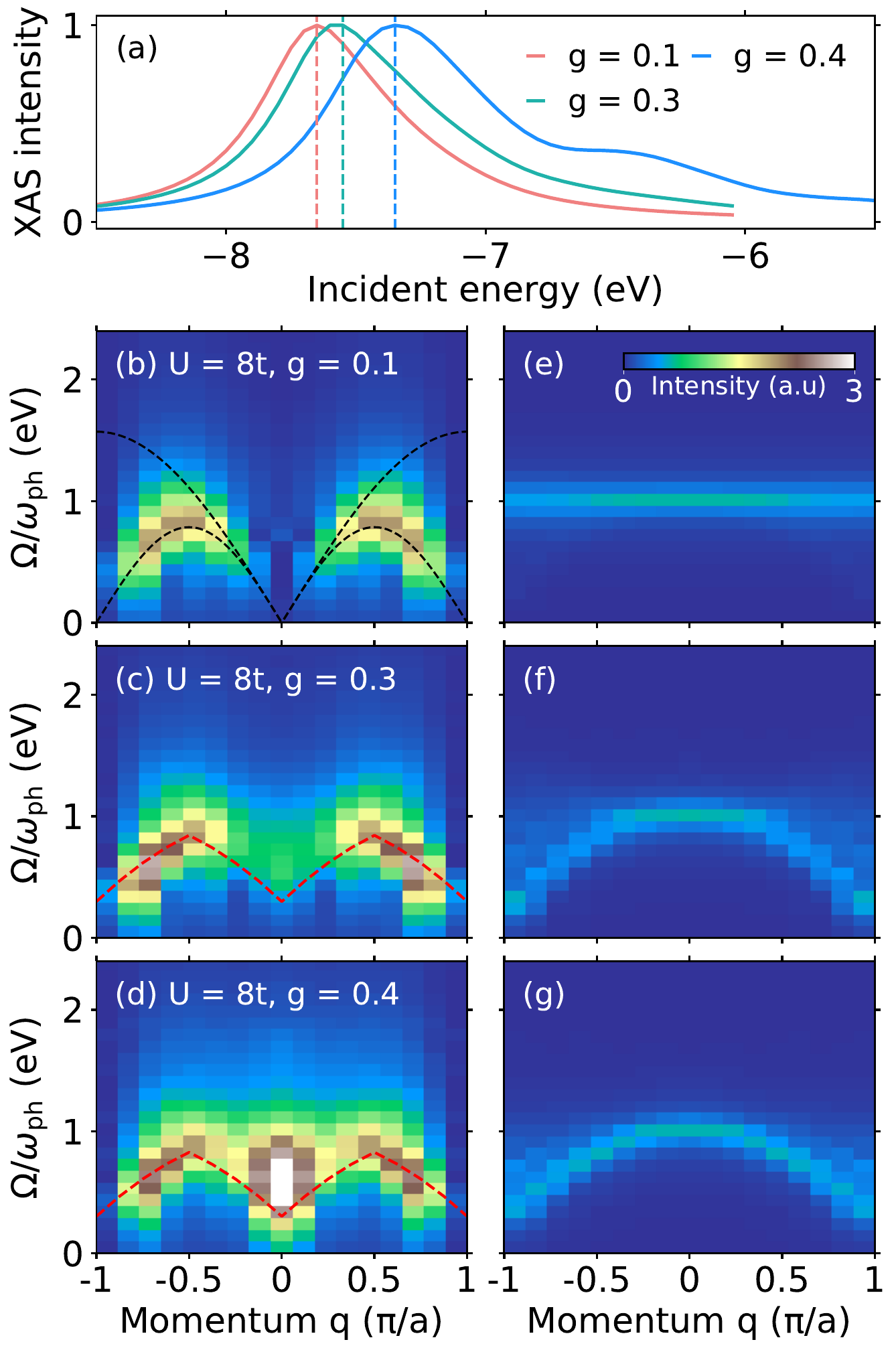}
    \caption{(a) Normalized \gls*{XAS} intensity for different values of $g$. The dashed lines indicate the incident energies in the subsequent \gls*{RIXS} calculations. (b-d) Plot spin-conserving \gls*{RIXS} spectra and (e-g) the corresponding phonon spectra functions for varying $g$ as indicated. The black and red dashed lines indicate the two-spinon and the lower boundary of the one-phonon + two-spinon continuum, respectively (see text). All results were obtained on half-filled $16$-site chains with $U=8t$, $\omega_\text{ph}=t$, $V_\mathrm{c} =-8t$, and $\Gamma/2=t/4$.}
    \label{L16_RIXS_spectra_U8_gchange}
\end{figure}
\noindent{\bf Results}. Figure~\ref{L16_RIXS_spectra_U8_gchange} shows \gls*{XAS} (Fig.~\ref{L16_RIXS_spectra_U8_gchange}a) and \gls*{RIXS} (Figs.~\ref{L16_RIXS_spectra_U8_gchange}b-d) results as a function of $g$ for fixed $U = 8t$. For reference, Figs.~\ref{L16_RIXS_spectra_U8_gchange}e-g plot the corresponding phonon spectral functions. Here, we focus on $g \le 0.4$ to ensure that the sign of the effective hopping does not change, which signals the breakdown of the linear approximation for the \gls*{SSH} model~\cite{Banerjee2023groundstate, Supplement}.

\begin{figure*}[t]
    \centering
    \includegraphics[width=0.95\textwidth]{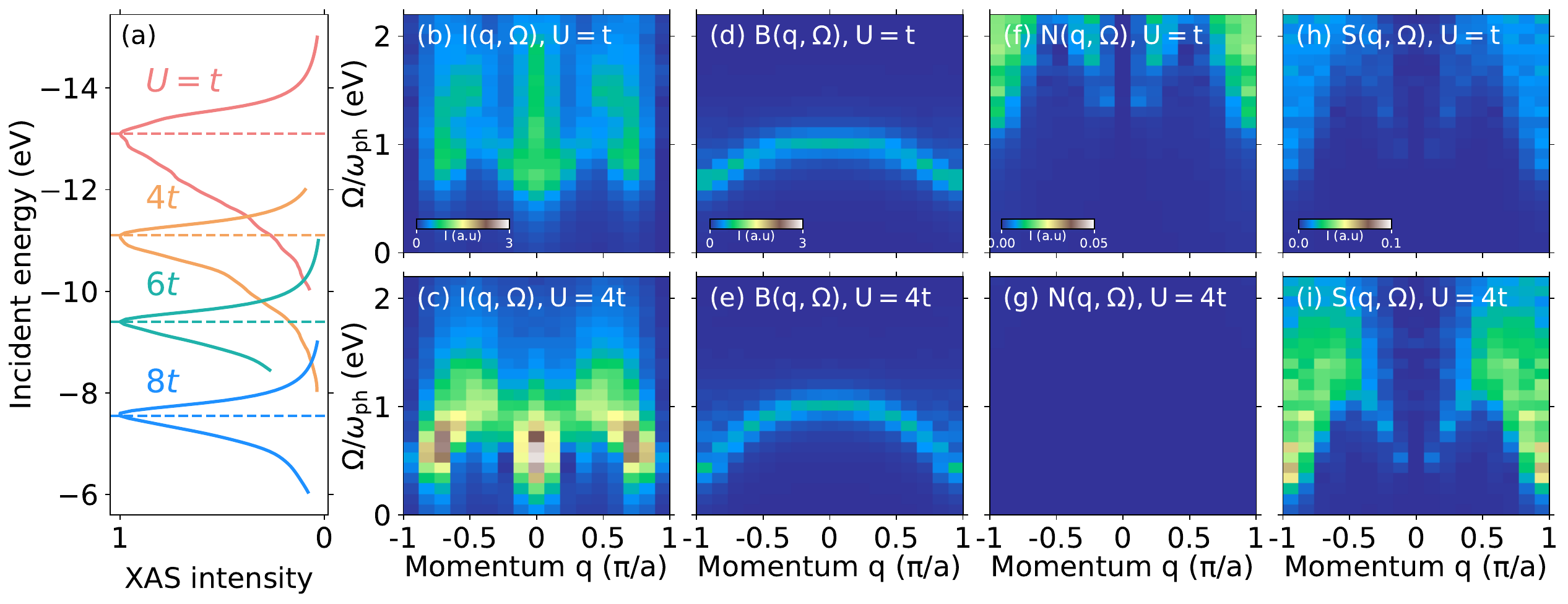}
    \caption{The evolution of the (a) XAS, (b),(c) \gls*{RIXS}, (d),(e) phonon spectral function, (f),(g) dynamical charge and (h),(i) spin structure factors as a function of $U$ as indicated. The spectra are computed with DMRG on $L=16$ site chains with $g = 0.3$, $\omega_\text{ph} = t$, $V_\mathrm{c}=-8t$ and $\Gamma/2=t/4$.}
    \label{L16_RIXS_spectra_g03_Uchange}
\end{figure*}

In the absence of \gls*{eph} couling ($g = 0$), the \gls*{XAS} has a single resonance corresponding to the well-screened final state in the plotted $\omega_\mathrm{in}$ range~\cite{Kourtis2012exact}. Increasing $g$ shifts this resonance to higher energy and creates weak Franck-Condon satellites at multiples of the phonon energy. The latter observation is similar to the predictions for the Holstein model~\cite{Thomas2024theory}, but the former is not. Increasing $g$ in the Holstein model shifts the well-screened resonance to lower incident energies due to the increased binding of the doublon. The behavior for the \gls*{SSH} interaction is reversed, indicating that it is less effective in localizing the excited core electron. 

We now turn to the \gls*{RIXS} spectra for $\omega_\mathrm{in}$ tuned to the main resonance in the \gls*{XAS} spectra. For weak coupling ($g = 0.1$), the spectra are dominated by the same $\Delta S = 0$ magnetic excitations that appear for the Hubbard model~\cite{Forte2011doping, Brink2011PRLspinexchangeDSF, Kumar2022unraveling}. Here, spinon excitations are concentrated along the lower boundary of the two-spinon continuum with vanishing weight at $q=0$ and $q=\pm\pi/a$. No clear phonon excitations are observed in the \gls*{RIXS} spectra and the phonon spectral function (Fig.~\ref{L16_RIXS_spectra_U8_gchange}e) tracks the bare phonon energy $\omega_\mathrm{ph} = t$. Increasing $g$ generates strong $Q = \pi/a$ \gls*{BOW} correlations, which soften the phonon dispersion at the zone boundary (Figs.~\ref{L16_RIXS_spectra_U8_gchange}f,g)~\cite{Banerjee2023groundstate, MalkarugeCosta2023comparative}. A new lattice-related excitation also appears in the \gls*{RIXS} spectra whose weight is concentrated at the zone center and grows with $g$. The details of this excitation differ significantly from the predictions obtained from the Holstein model, where the one-phonon excitations probed by \gls*{RIXS} closely track the renormalized phonon dispersion~\cite{Thomas2024theory}. 
Instead, the phonon-related excitations observed here in the \gls*{RIXS} spectra are redshifted at $q = 0$ and appear at an energy comparable to the $q = \pi/a $ soft phonon mode. We also cannot resolve soft phonon modes in the \gls*{RIXS} spectra at $q = \pm \pi/a$, although they may overlap with the magnetic excitations at these momenta. Another curious observation is that the intensity of the phonon features is largest at $q = 0$ while the \gls*{eph} matrix element vanishes as $q \rightarrow 0$ for our model~\cite{Li2011perturbation}. This observation counters the general expectation that the intensity of the phonons observed in \gls*{RIXS} track the momentum-dependence of the microscopic \gls*{eph} coupling~\cite{Ament2011determining, Devereaux2016directly}. 

\begin{figure}[t]
    \centering
    \includegraphics[width=\columnwidth]{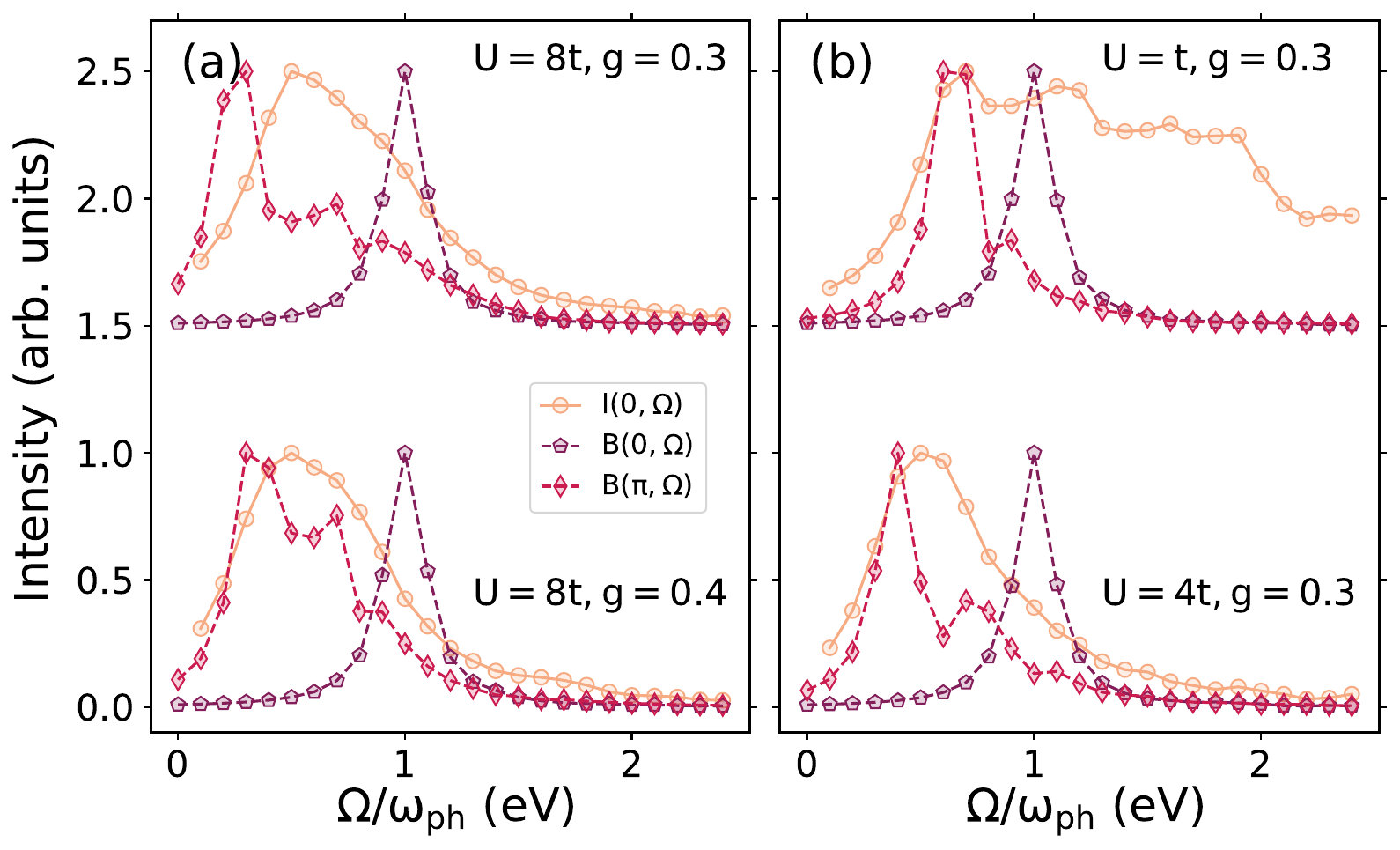}
    \caption{A comparison between the $q = 0$ spin-conserving RIXS spectra and the phonon spectral function $[B(q,\Omega)]$ at $q=0$ and $q=\pi/a$. Panel (a) shows results for fixed $U=8t$ with different $g$ while panel (b) shows results for $g=0.3$ and different $U$, as indicated. The spectra are computed with DMRG in a $L=16$ site chain with $\omega_\text{ph} = t$, $V_\mathrm{c}=-8t$ and $\Gamma/2=t/4$.}
    \label{L=16_RIXS_first_Peak}
\end{figure}

Figure~\ref{L16_RIXS_spectra_g03_Uchange} examines the $U$-dependence of the spectra for fixed $g = 0.3$. Decreasing $U$ shifts the well-screened \gls*{XAS} resonance to lower incident energies, reflecting the change in the potential energy of the intermediate state doublon. It also skews the spectra to higher incident energies due to a combination of Franck-Condon broadening and decreased electron correlations~\cite{Thomas2024theory}. The \gls*{RIXS} spectra and corresponding phonon, charge, and spin correlation functions are shown in the remaining panels, as indicated. For $U=t$ [Fig.~\ref{L16_RIXS_spectra_g03_Uchange}b], the \gls*{RIXS} spectra have contributions from charge, two-spinon, and lattice-related excitations. The spinons appear as the arcing intensity at large $q$ and extend to higher energies due to the increased exchange coupling. Conversely, the charge excitations are concentrated at $q=0$ where they disperse rapidly to merge with the phonon excitations at low energy. We note that this distinction is somewhat arbitrary as the lattice and charge excitations are strongly mixed for this value of $U$, as described below. Increasing $U$ Mott-gaps the charge fluctuations at high energy, which concentrates intensity of the $q \approx 0$ lattice-related excitations and spinon excitations at lower energy, bringing the spectra more in line with the large-$U$ results shown in Fig.~\ref{L16_RIXS_spectra_U8_gchange}. 

As with Fig.~\ref{L16_RIXS_spectra_U8_gchange}, none of the \gls*{RIXS} spectra shown in Fig.~\ref{L16_RIXS_spectra_g03_Uchange} 
exhibit any excitation that tracks the renormalized phonon dispersion. To examine the relevant excitations further, Fig.~\ref{L=16_RIXS_first_Peak} compares the energy of the $q=0$ \gls*{RIXS} spectra to the phonon spectral functions at $q=0$ and $\pi/a$ for selected values of $U$ and $g$. In all cases, the onset of the \gls*{RIXS} spectra is bound by the energy of the $q=\pi/a$ phonon mode rather than the $q = 0$ mode. This result radically differs from the Holstein model predictions, where the lattice excitations probed by \gls*{RIXS} coincide with the renormalized phonon energies~\cite{Thomas2024theory}. We primarily attribute this difference to the nature of the \gls*{SSH} interaction, which modulates the electronic hopping rather than the on-site energy. This aspect makes it impossible to excite the lattice without also exciting the electronic subsystem. For example, in the Mott insulating regime, exciting the \gls*{SSH} phonons necessarily modulates the exchange interaction in the intermediate state of the \gls*{RIXS} process. Similarly, in the metallic regime, exciting the lattice will also naturally lead to density modulations via the changes in the effective hopping.  The lattice-related excitations observed here are thus likely multi-particle excitations where the energy and momentum of the scattered photon is distributed between the phonons and other spin and charge excitations.  

To confirm the multi-particle nature of this excitation, the red dashed lines in Fig.~\ref{L16_RIXS_spectra_U8_gchange} plot the lower boundary of the one-phonon plus two-spinon continuum, which we derived from the renormalized phonon dispersion and the lower boundary of the two-spinon continuum (see Fig.~\ref{L16_RIXS_spectra_U8_gchange}b). The excitations probed by \gls*{RIXS} near $q=0$ are fully bound by this energy scale with the minimum energy for the $q=0$ \gls*{RIXS} excitations being set by the energy of a $q = \pm\pi/a$ phonon mode together with a $q = \mp\pi/a$ two-spinon excitation. These phase space considerations thus strongly imply that the new excitations appearing when $g\ne0$ are not pure phonon excitations as in the Holstein model but rather multi-particle excitations involving a phonon plus two spinons or phonons plus density fluctuations. This scenario also explains why the intensity of the excitations peaked at the zone center, even though the \gls*{SSH} model's \gls*{eph} matrix element $g(k,q)$ is largest at $q = \pm\pi/a$~\cite{Li2011perturbation}. 

\begin{figure}[t]
    \centering
    \includegraphics[width=\columnwidth]{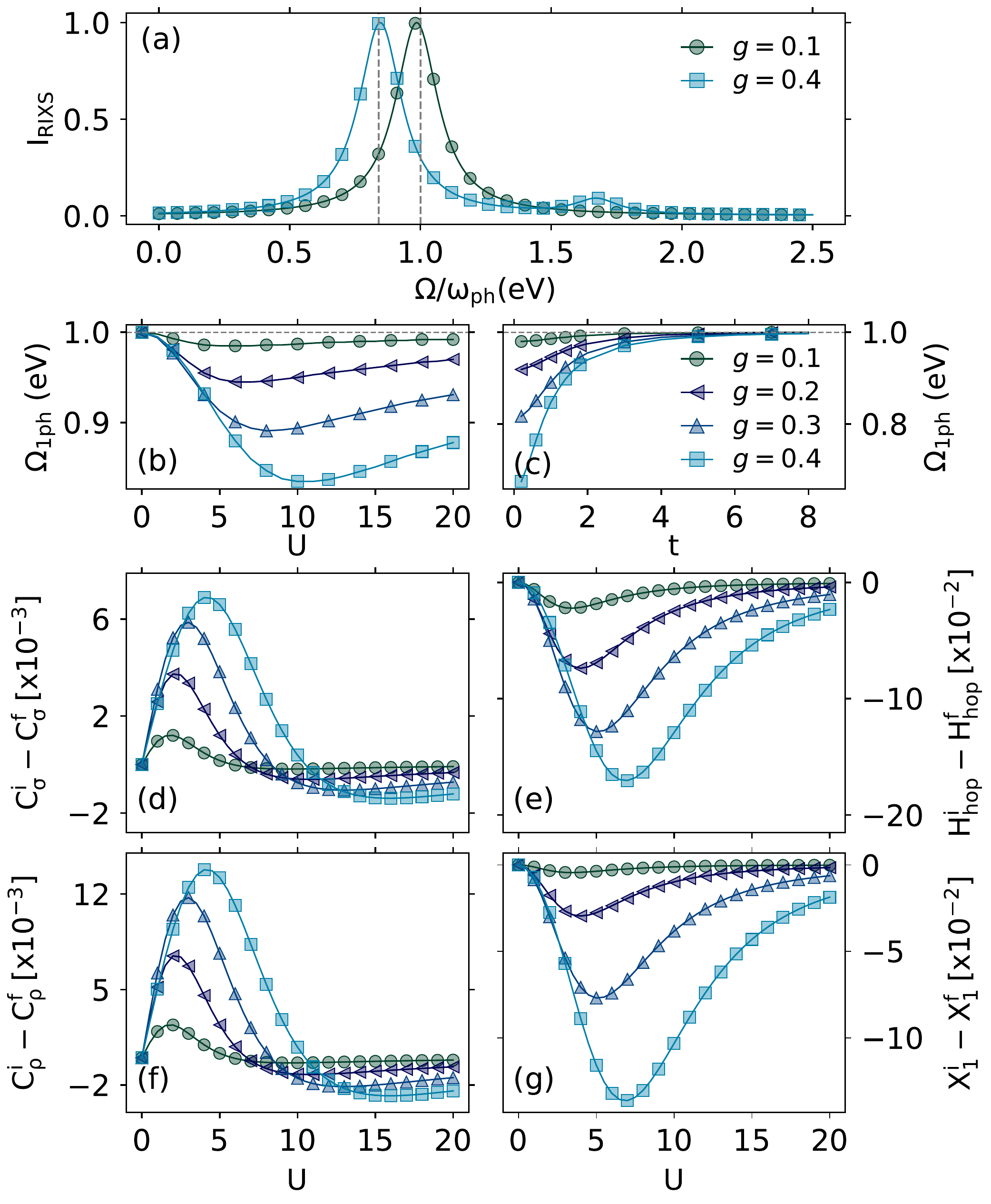}
    \caption{(a) Normalized RIXS spectra at $q=0$ calculated with Hubbard-SSH dimer ED calculation for $U=8t$ and varying $g$ as indicated. Dashed lines show the energy loss of the first phonon peak. (b) Energy loss of the one-phonon RIXS peak as a function of $U$ for fixed $t=1$  and (c) $t$ for fixed $U=8$ with varying $e$-ph coupling $g$. Panels (d) - (g) show results for the difference of spin correlation, the expectation of electron hopping, double occupancy, and lattice distortion between the initial and the final state (one-phonon peak), as a function of $U$ for fixed $t=1$, with varying $g$. All results are obtained with ED on a \gls*{HSSH} dimer for $\omega_\text{ph}=t$, $V_\mathrm{c} = -8t$, and $\Gamma/2=t/4$.}
    \label{2site_ED_correlations}
\end{figure}

To confirm this scenario, Fig.~\ref{2site_ED_correlations} presents an \gls*{ED} analysis of the \gls*{RIXS} spectra and wave functions of a \gls*{HSSH} dimer. Fig.~\ref{2site_ED_correlations}a shows a representative normalized \gls*{RIXS} spectra at $q = 0$ with $U = 8t$ and $g = 0.1,~0.4$. We extract the energy loss position $\Omega_{1ph}$ of the low-energy excitation directly from these spectra as indicated by the dashed line, and track its evolution as a function of $U$ and $t$, as shown in Figs.~\ref{2site_ED_correlations}b and \ref{2site_ED_correlations}c, respectively. The excitation energy redshifts for any value of $U\ne 0$ by an amount that depends on $g$ and $t$. Crucially, the magnitude of the shift is nonmonotonic as a function of $U$ and resembles the evolution of magnetic correlations in the Hubbard model~\cite{Staudt2000phase}. Conversely, increasing $t$ (Fig.~\ref{2site_ED_correlations}c) drives $\Omega_{1ph}$ toward the bare phonon energy. 

We also examined the change of several correlation functions between the initial ($i$) and final ($f$) states of the \gls*{RIXS} process. These include the spin-spin correlation $C_{\sigma}^{m} = \bra{m}S_1^z S_2^z\ket{m}$, the double occupancy $C_{\rho}^{m} = \bra{m}n_{1,\uparrow}n_{1,\downarrow}\ket{m}$, the single particle hopping $H_\text{hop}^{m} = \bra{m}c_{1,\sigma}^{\dagger}c^{\phantom\dagger}_{2,\sigma} + \text{h.c.}\ket{m}$, and the average displacement $X_1^{m} = \bra{m}(b_1^{\dagger} + b_1^{\phantom\dagger})\ket{m}$, where $\ket{m}$ is an eigenstate of the system. The evolution of these quantities as a function of $U$ is shown in Figs.~\ref{2site_ED_correlations}d-g, where we consider the difference in each quantity for the ground state and first excited state visible in the \gls*{RIXS} spectra. Each shows significant changes between the scattering process's initial and final states. This behavior is in stark contrast to Holstein \gls*{eph} interaction, where there is no discernible modification to the electronic sector for the first phonon excitation probed by \gls*{RIXS}~\cite{Supplement}. These \gls*{ED} results further support the idea that exciting the lattice via an \gls*{SSH} \gls*{eph} interaction inevitably also excites the system's magnetic and electronic sectors.\\

\noindent{\bf Discussion}. We have studied the \gls*{RIXS} spectra for a half-filled \gls*{HSSH} model using \gls*{DMRG} and \gls*{ED} methods. We found that the \gls*{SSH} interaction produces low-energy multi-particle excitations that are not purely phononic in character. In particular, when the system has a strong exchange interaction, the low-energy excitations combine one-phonon plus two-spinon excitations, which appear at energies well below the $q = 0$ phonon modes. They actually track the softened $q =\pi$ phonon mode, which is driven by the SSH \gls*{eph} coupling tendency to dimerize the system at half-filling ($\pi=2k_F$). It would be interesting to confirm this scenario away from half-filling in future studies. Moreover, the intensity of these excitations does not track the $q$-dependence of the bare \gls*{eph} coupling constant~\cite{Ament2011resonant}. While these results complicate the interpretation of experiments, they also suggest an experimental protocol for identifying the presence of \gls*{SSH} interactions in strongly correlated materials. Specifically, we expect that the energy of the zone center lattice excitations measured in a \gls*{RIXS} experiment will be lower than the values measured with optical or Raman spectroscopy if they arise from an \gls*{SSH}-like interaction. \\

\noindent{\bf Acknowledgments}. We thank M. Berciu, B. Cohen-Stead, and N. C. Plumb for useful discussions and comments. This study is supported by the National Science Foundation under Grant No. DMR-1842056 and used resources from the Oak Ridge Leadership Computing Facility, which is a DOE Office of Science User Facility supported under Contract No. DE-AC05-00OR22725. A. N. acknowledges the support of the Canada First Research Excellence Fund. 

\bibliography{references}
\end{document}


\preprint{}
\title{Supplementary Material for ``Identifying and quantifying Su-Schrieffer-Heeger-like interactions using RIXS''}

\author{Debshikha Banerjee\,\orcidlink{0009-0001-2925-9724}}
\affiliation{Department of Physics and Astronomy, The University of Tennessee, Knoxville, Tennessee 37996, USA}
\affiliation{Institute for Advanced Materials and Manufacturing, University of Tennessee, Knoxville, Tennessee 37996, USA\looseness=-1}
\author{Jinu Thomas\,\orcidlink{0000-0003-4818-6660}}
\affiliation{Department of Physics and Astronomy, The University of Tennessee, 
Knoxville, Tennessee 37996, USA}
\affiliation{Institute for Advanced Materials and Manufacturing, University of Tennessee, Knoxville, Tennessee 37996, USA\looseness=-1}
\author{Alberto Nocera\,\orcidlink{0000-0001-9722-6388}}
\affiliation{Stewart Blusson Quantum Matter Institute, University of British Columbia, Vancouver, British Columbia, Canada
V6T 1Z4}
\affiliation{Department of Physics Astronomy, University of British Columbia, Vancouver, British Columbia, Canada V6T 1Z1}
\author{Steven Johnston\,\orcidlink{0000-0002-2343-0113}}
\affiliation{Department of Physics and Astronomy, The University of Tennessee, Knoxville, Tennessee 37996, USA}
\affiliation{Institute for Advanced Materials and Manufacturing, University of Tennessee, Knoxville, Tennessee 37996, USA\looseness=-1}

\date{\today}

\maketitle


\section{Calculation of XAS and other dynamical correlation functions}
This section provides details about how we calculate the \gls*{XAS} spectra, phonon spectral function $B(q,\omega)$, and dynamical spin $S(q,\omega)$ and charge $N(q,\omega)$ structure factors. 

The \gls*{XAS} spectra are calculated using Fermi's golden rule 
\begin{align}
\label{eq: XAS_intensity}
I_{\mathrm{XAS}}(\omega_{\mathrm{in}})&\propto  
\sum_n \left\vert\sum_{\sigma}\langle n|D_{j,\sigma}|i\rangle\right\vert^2 
\delta(E_{i}-E_{n}+\omega_{\mathrm{in}}) 
\propto - \mathrm{Im} 
\sum_{n}\frac{\left\vert\sum_{\sigma}\langle n|D_{j,\sigma}|i\rangle\right\vert^2}{E_{i}-E_{n}+\omega_{\mathrm{in}}+\mathrm{i}\Gamma/2}.
\end{align}
In the last equality, we have set the energy broadening of the \gls*{XAS} spectra to equal the inverse core-hole lifetime. 

The phonon spectral function and the dynamical spin and charge structure factors are obtained from the Fourier transform~\cite{nocera2016spectral} of their relevant real-space correlation functions 
\begin{align}
    B_{j,c}(\Omega)&= -\frac{1}{\pi} \operatorname{Im} [\bra{i}\Tilde{X}_j G\Tilde{X}_c \ket{i}]\label{eq:Brw}, \\
    N_{j,c}(\Omega)& = -\frac{1}{\pi} \operatorname{Im} [\bra{i}\Tilde{n}_j G\Tilde{n}_c \ket{i}],~\text{and}\label{eq:Nrw}\\
    S_{j,c}(\Omega)& = -\frac{1}{\pi} \operatorname{Im} [\bra{i}S^z_j G S^z_c \ket{i}]. \label{eq:Srw}
\end{align}
Here, $c$ is the center site of the chain, $\ket{i}$ is the ground state of the system with energy $E_i$, $\Tilde{X}_j = (\hat{X}_j-\langle \hat{X}_j \rangle)$, $\Tilde{n}_j = (\hat{n}_j-\langle \hat{n}_j \rangle)$, and 
\begin{align}\label{eq:Gfinal}
G &=  \frac{1}{E_{i}-\hat{\mathcal{H}}+\Omega+\mathrm{i}\eta}
\end{align} 
is the electron Green's function. We compute Eqs.~\eqref{eq: XAS_intensity}-\eqref{eq:Srw} using DMRG++ with center-site approximation and the correction vector algorithm~\cite{Kuhner1999, Jeckelmann2002, nocera2016spectral}. 

\newpage
\section{Critical electron-phonon coupling for with hopping inversion}
Within the linear approximation, the effective hopping integral for the \gls*{SSH} model is  $t_{\mathrm{eff}} \approx -t + g \langle X_i - X_{i+1} \rangle$. If the  \gls*{eph} coupling is strong, the average lattice displacements can become large enough to change the sign of $t_{\mathrm{eff}}$. When this occurs, the system is unstable toward dimerization and the model becomes unphysical~\cite{Banerjee2023groundstate}. 

\begin{figure}[h]
    \centering
    \includegraphics[width=0.4\textwidth]{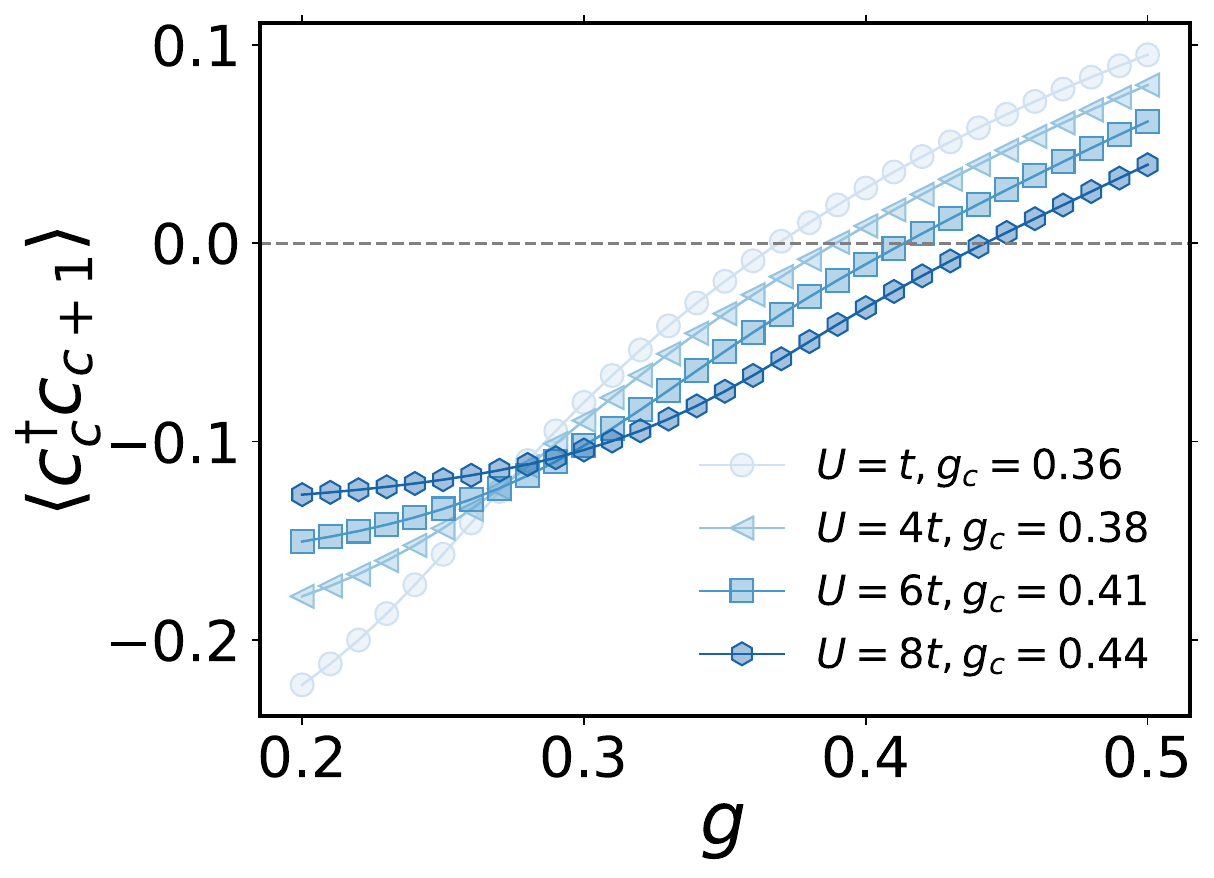}
    \caption{Ground state expectation values of single-particle hopping $\langle H_{\mathrm{hop}} \rangle$ for the \gls*{HSSH} model, calculated with DMRG on an $L=16$ site 1D half-filled chain. The expectation values were calculated for hopping between the center site and its neighbor as a function of $g$ for different $U$ values as indicated. These expectation values become positive when the effective hopping changes sign, as discussed in Ref.~\cite{Banerjee2023groundstate}. For $g > g_c$, the linear \gls*{SSH} model is no longer valid.} 
    \label{fig: critical_coupling}
\end{figure}

The critical coupling $g_\mathrm{c}$ for sign inversion can be determined by monitoring the average value of the single-particle hopping $\langle H_\text{hop} \rangle = \langle c^{\dagger}_c c^{\phantom\dagger}_{c+1} \rangle$ at the chain's center. To this end, Fig.~\ref{fig: critical_coupling} plots the ground-state expectation values of the single-particle hopping for different values of $U$. These results were obtained using \gls*{DMRG} applied to a half-filled \gls*{HSSH} chain with $L = 16$ sites and $\omega_\text{ph}=t$, as in the main text. Increasing $U$ shifts the critical coupling $g_\mathrm{c}$ to larger values.

\newpage
\section{RIXS spectra in spin-conserving and non-spin-conserving channels}
\begin{figure}[h]
    \centering
    \includegraphics[width=\columnwidth]{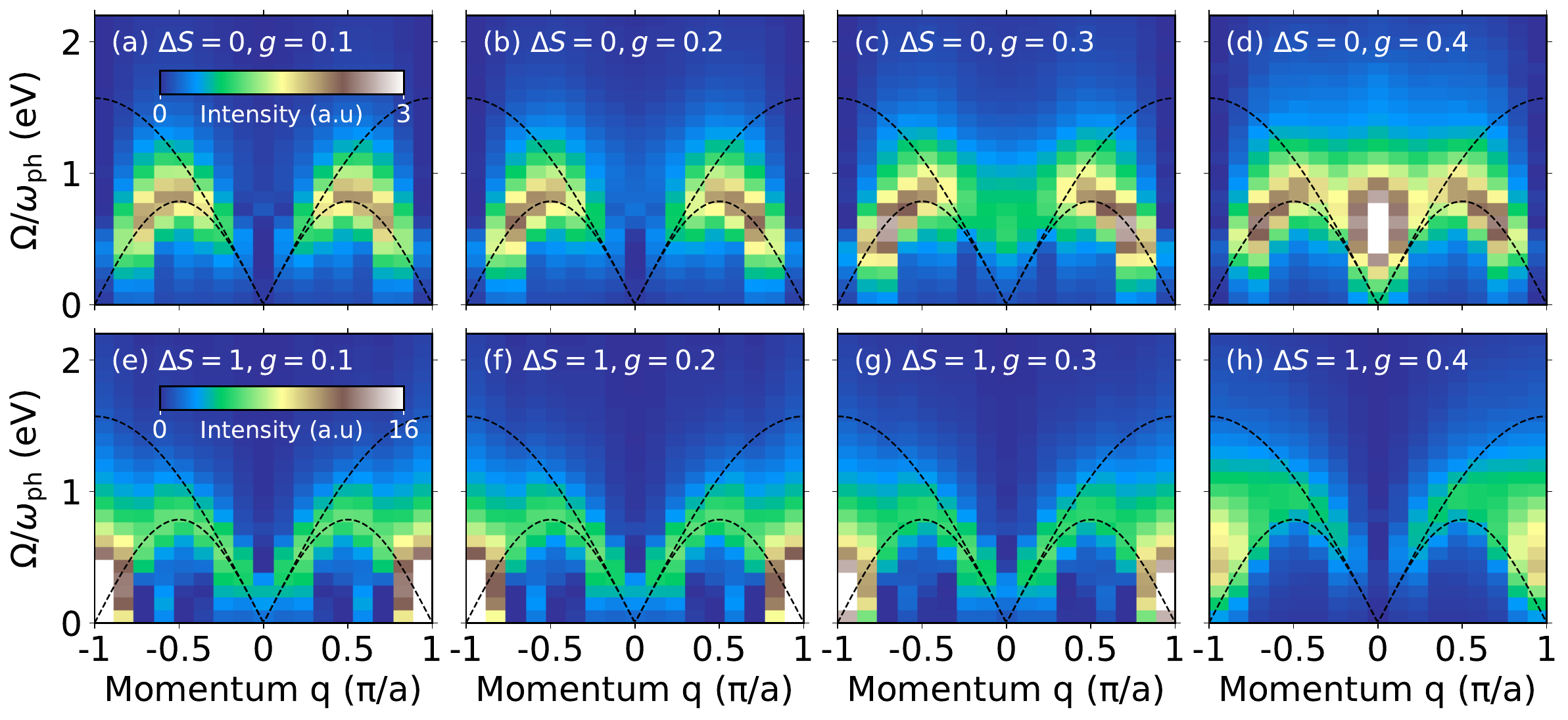}
    \caption{The top and bottom rows show RIXS spectra in the spin-conserving ($\Delta S = 0$, panels a-d) and non-spin-conserving ($\Delta S = 1$, panels e-h) channels, respectively. All results were obtained on half-filled $L=16$ site chains with $U=8t$, $\omega_\text{ph}=t$, $\Gamma/2=t/4$, $V_\mathrm{c}=-8t$ and varying $g$ as indicated. The black dashed lines show the lower and upper boundaries of the two-spinon continuum assuming $J=4t^2/U$, as expected for the Hubbard model in the strong coupling limit.}
    \label{fig:RIXS_NSC_SC_U8}
\end{figure}

Figure~\ref{fig:RIXS_NSC_SC_U8} shows \gls*{RIXS} spectra in the spin-conserving ($\Delta S=0$, Figs.~\ref{fig:RIXS_NSC_SC_U8}a-d) and non-spin-conserving ($\Delta S=1$, Figs.~\ref{fig:RIXS_NSC_SC_U8}e-h) channels for $U=8t$, $\omega_\text{ph}=t$, $\Gamma/2=t/4$, $V_\mathrm{c}=-8t$, and increasing \gls*{eph} coupling $g$, as indicated. 

The spin-conserving \gls*{RIXS} spectra have contributions from magnetic excitations generated from double spin-flips that produce anti-parallel two-spinon excitations. The spectral weight for these excitations peak at $q=\pi/2a$ and vanish at $q=0$ and $q=\pi/a$ and can be well approximated by the dynamical spin exchange structure factor~\cite{Forte2011doping, Brink2011PRLspinexchangeDSF}. This channel also has contributions from multi-particle excitations generated by the \gls*{eph} coupling, as discussed in the main text. The non-spin-conserving channel, in contrast, is dominated by two-spinon excitations that are generated by direct spin-flip excitations. This scattering channel is analogous to the spin-flip \gls*{INS} and the \gls*{RIXS} spectra in this channel closely resemble the dynamical spin structure factor for a Hubbard chain in the strong coupling limit~\cite{nocera2016spectral, bhaseen2005itinerancy, Brink2011PRLspinexchangeDSF}. Crucially, we find no indications of additional lattice excitations in this channel, which justifies our focus in the main text. 

\newpage
\section{$V_\mathrm{c}$ dependence of RIXS spectra}
\begin{figure}[h]
    \centering
    \includegraphics[width=\columnwidth]{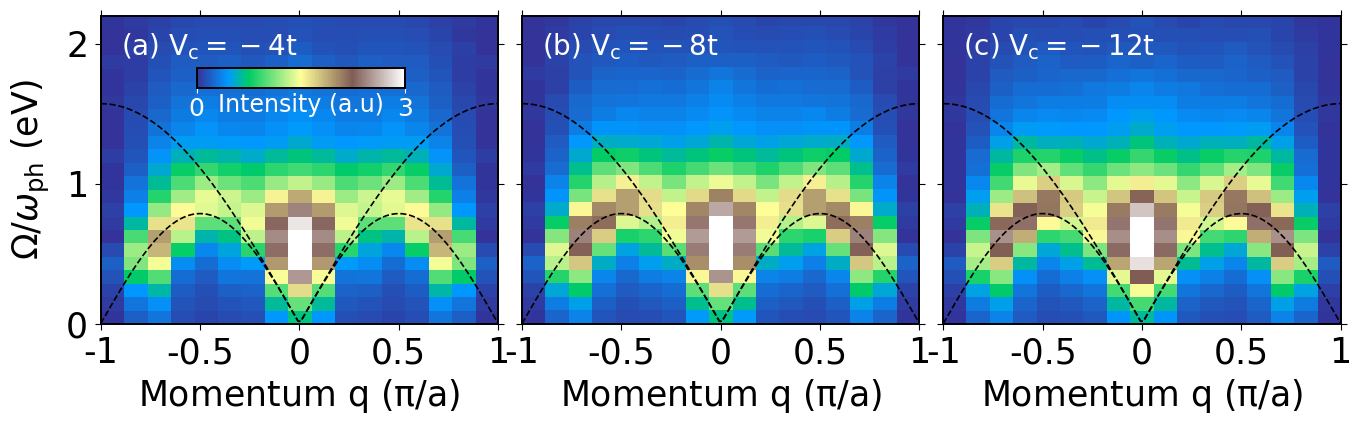}
    \caption{RIXS spectra in the spin-conserving channel for $U=8t$, $\omega_\text{ph}=t$, $\Gamma/2=t/4$, $g=0.4$ and varying $V_\mathrm{c}$, as indicated. The black dashed lines show the lower and upper boundaries of the two-spinon continuum, assuming $J=4t^2/U$ for the Hubbard model. All spectra have been plotted on the same color scale, which is shown in panel a.}
    \label{fig:RIXS_Vc_vary_U8_g04}
\end{figure}

Figure~\ref{fig:RIXS_Vc_vary_U8_g04} shows \gls*{RIXS} spectra in the spin-conserving channel as a function of the core-hole potential $V_\mathrm{c}$, as indicated. These results were obtained using \gls*{DMRG} on an $L = 16$ site chain with fixed  $U=8t$, $\omega_\text{ph}=t$, $\Gamma/2=t/4$, and $g=0.4$. In the \gls*{HH} model, a stronger core-hole potential tends to localize the intermediate state doublon and can often generate more intense phonon excitations in the spectra~\cite{Thomas2024theory}. Here, however, we do not see a considerable increase in the intensity of the phonon peak at $q=0$ as $V_\mathrm{c}$ changes from $-4t$ to $-12t$ but we do observe an overall enhancement of the spinon excitations relative to the phonon excitations. We attribute this behavior to the spin-phonon coupling inherent to the \gls*{SSH} interaction. We also note that the location of the phonon-related multi-particle excitations remains unchanged as the magnitude of $V_\mathrm{c}$ increases.

\newpage
\section{Comparison of the RIXS spectra, phonon spectral function, and dynamical charge and spin structure factors}
\begin{figure}[h]
    \centering
    \includegraphics[width=1.0\textwidth]{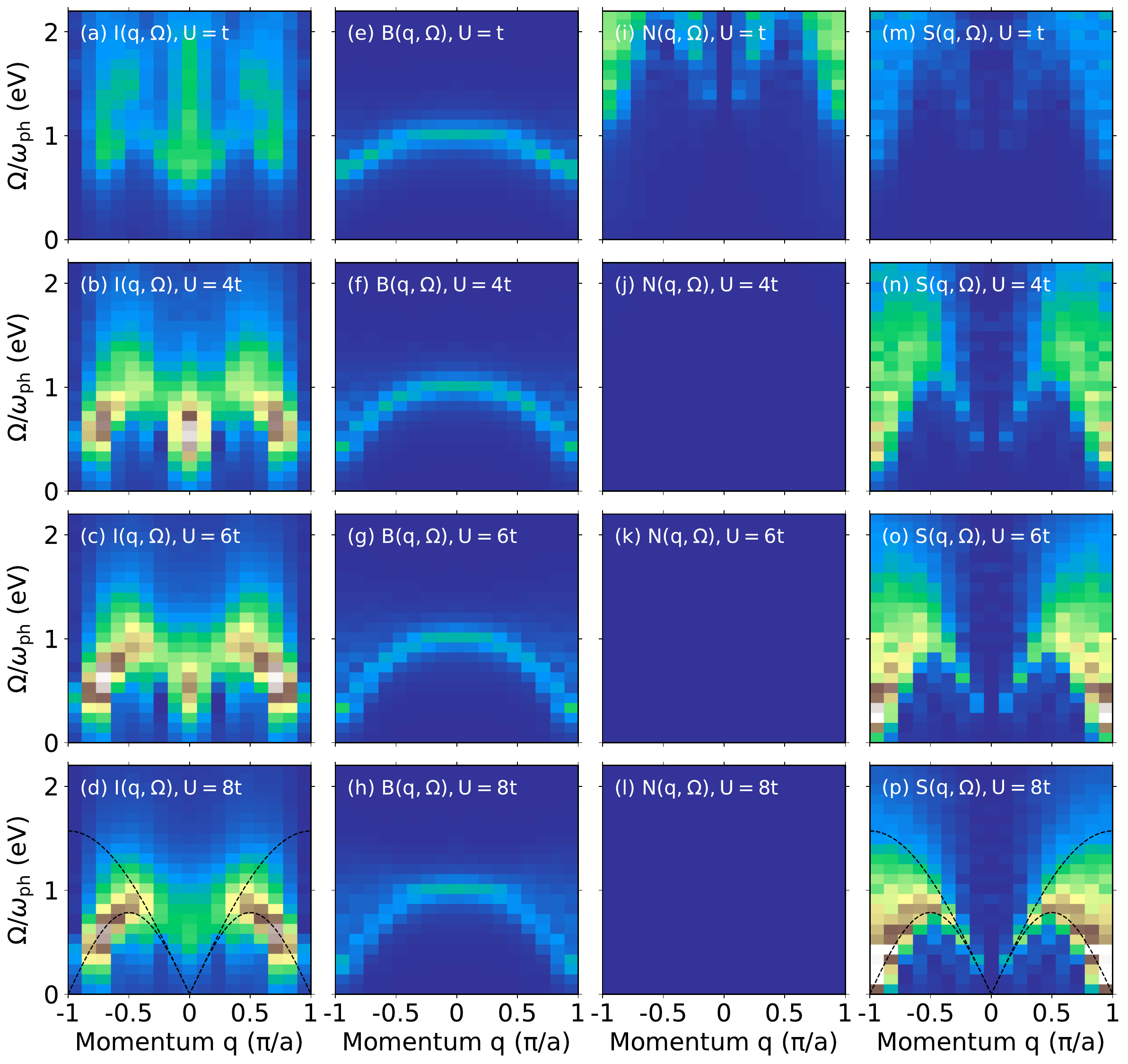}
    \caption{(a)-(d) \gls*{RIXS} spectra in the spin-conserving channel 
    for varying values of $U$, as indicated. The second through fourth columns show the corresponding phonon spectral functions $B(q,\omega)$ [panels (e)-(h)], dynamical charge structure factors $N(q,\omega)$ [panels (i)-(l)], and dynamical spin structure factors $S(q,\omega)$ [panels (m)-(p)]. All results were obtained on $L = 16$ site chains with fixed $g=0.3$, $\omega_\text{ph} = t$, $V_\mathrm{c}=-8t$, and $\Gamma/2=t/4$. 
    }
    \label{fig:RIXS_Bqw_Nqw_Sqw_g03_varyU}
\end{figure}

Figure~\ref{fig:RIXS_Bqw_Nqw_Sqw_g03_varyU} compares the spin-conserving \gls*{RIXS} spectra to the phonon spectral function [$B(q,\Omega)$] and dynamical charge [$N(q,\Omega)$] and spin [$S(q,\Omega)$] structure factors for $g=0.3$, $V_\mathrm{c}=-8t$, $\Gamma/2=t/4$ and varying $U$ as indicated. The results for $U = t$ and $4t$ are identical to those shown in Fig. 2 of the main text. 

\newpage
\section{Convergence of our DMRG results}
\begin{figure}[h]
    \centering
    \includegraphics[width=\columnwidth]{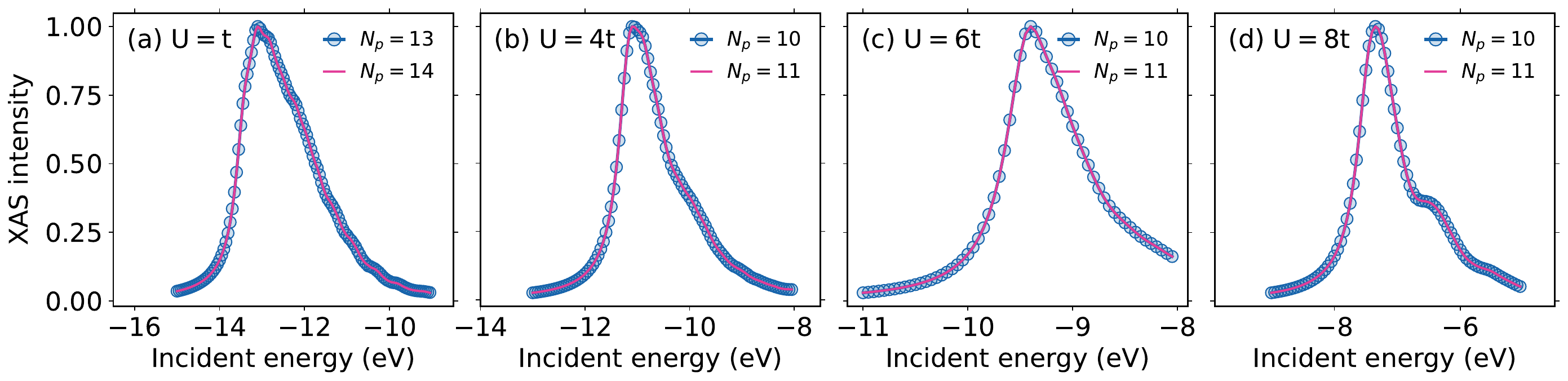}
    \caption{Convergence of XAS intensity with respect to the size of the local phonon Hilbert space $N_p$. Results are shown for (a) $U=t$, $g=0.3$, (b) $U=4t$, $g=0.3$, (c) $U=6t$, $g=0.4$, and (d) $U=8t$, $g=0.4$. All results are calculated on an $L=16$ site \gls*{1D} chain with $\omega_\text{ph}=t$, $V_\text{c}=-8t$, and $\Gamma/2=t/4$.}
    \label{fig:XAS_conv_Np}
\end{figure}

This section provides the results of several convergence tests performed on our \gls*{DMRG} simulations. Fig.~\ref{fig:XAS_conv_Np} shows the convergence of \gls*{XAS} spectra with respect to the size of the local phonon Hilbert space $N_p$. Here, we show results for $g=0.3$ with (a) $U=t$ and (b) $U=4t$ and for $g=0.4$ with (c) $U=6t$ and (d) $U=8t$. In all cases, our \gls*{XAS} results are well converged with respect to the number of phonon modes kept per site. 

\begin{figure}[h]
    \centering
    \includegraphics[width=\columnwidth]{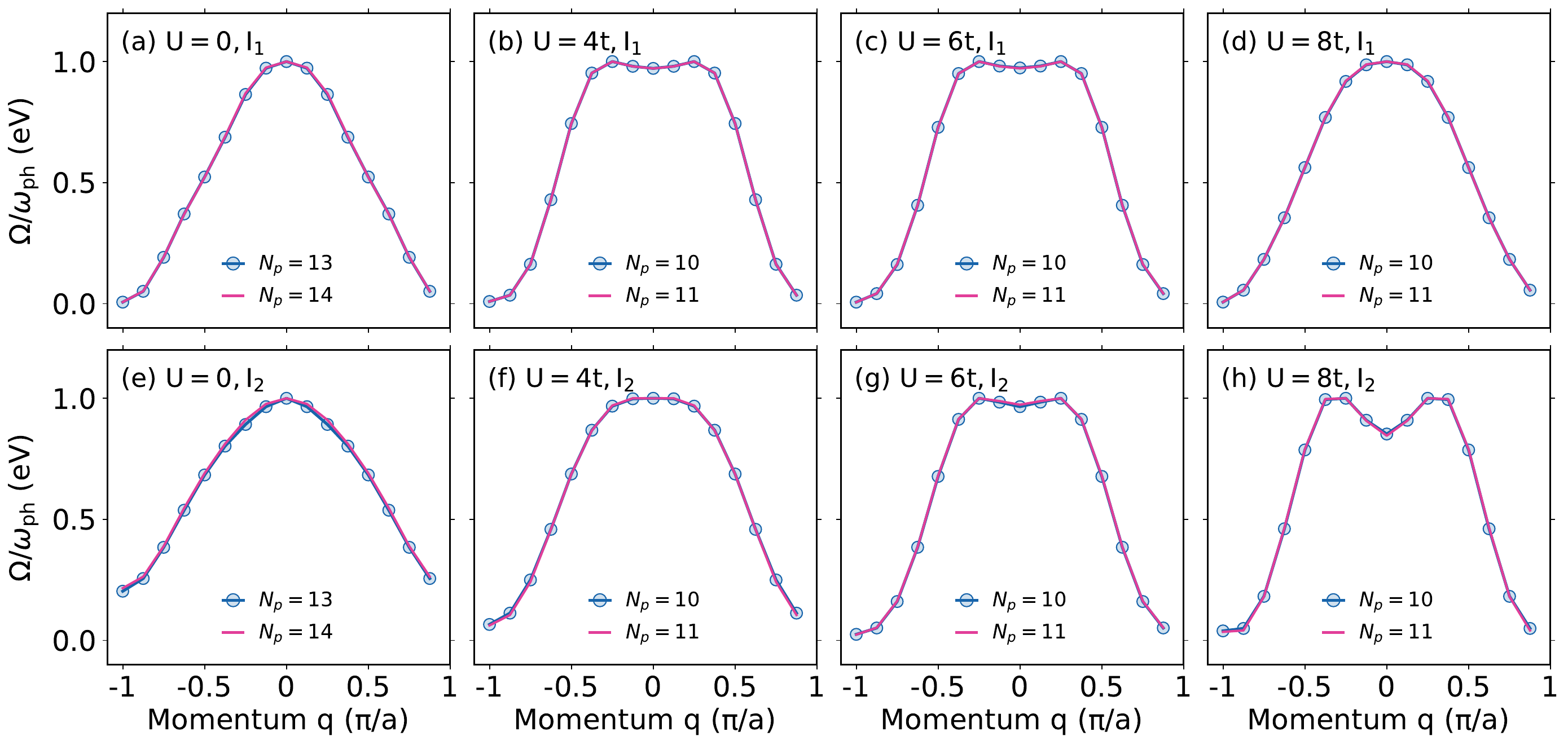}
    \caption{Convergence of RIXS intensity at the one- [$I_1$, panels (a)-(d)] and two-phonon [$I_2$, panels (e)-(h)] peaks as a function of the number of phonon modes per site ($N_p$) for varying $U$, as indicated.  All results were calculated on an $L=16$ site \gls*{1D} chain with $g=0.5$, $\omega_\text{ph}=t$, $V_\text{c}=-8t$, and $\Gamma/2=t/4$.}
    \label{fig:RIXS_conv_Np}
\end{figure}

Figure~\ref{fig:RIXS_conv_Np} assesses the convergence of the \gls*{RIXS} spectra with respect to $N_p$ for $g=0.5$. Here, we considered a case with $g > g_c$ for two reasons. First, this value necessitates a large local phonon space to describe the coherent lattice distortions associated with the lattice dimerization~\cite{Nocera2021bipolaron}. We expect the spectra at $g < g_c$ to converge well for much smaller values of $N_p$. Second, we can more easily extract the locations of the first and second phonon excitations when the system has dimerized. Figs.~\ref{fig:RIXS_conv_Np}a-d and \ref{fig:RIXS_conv_Np}e-h plot the \gls*{RIXS} intensity along the one- and two-phonon peaks, respective, for varying $U$, as indicated in each panel. All plots show very little change in the spectra upon increasing $N_p$ from 13 to 14 (for $U=0$) and 10 to 11 (for $U=4t$ to $8t$), indicating that our  \gls*{RIXS} spectra are well converged.

\newpage
\section{Exact diagonalization and additional results for the Hubbard-Holstein model}

\begin{figure}[h]
    \centering
    \includegraphics[width=0.5\textwidth]{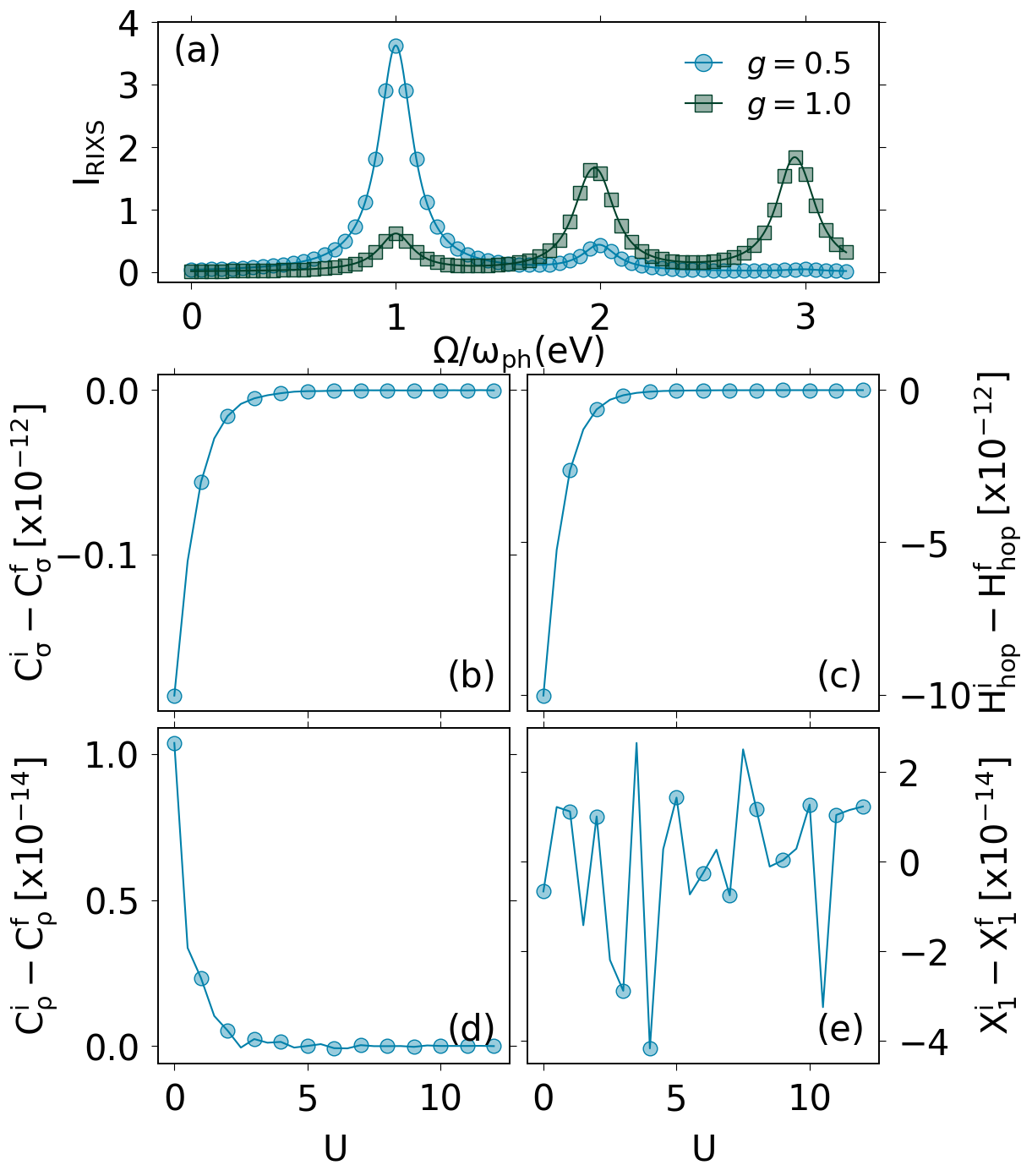}
    \caption{(a) RIXS spectra at $q=0$ calculated with Hubbard-Holstein dimer ED calculation for $U=8t$ and varying $g$ as indicated. Panels (b) - (e) show results for the difference of spin correlation, the expectation of electron hopping, double occupancy, and lattice distortion between the initial and the final state (one-phonon peak), as a function of $U$ for $g=0.5$. For all the results $\omega_\text{ph}=t$, $V_\text{c}=-8t$, $\Gamma/2=t/4$ are kept fixed.}
    \label{fig:2site_ED_Holstein}
\end{figure}

The main text presents \gls*{ED} results for a \gls*{HSSH} dimer. Here, Fig.~\ref{fig:2site_ED_Holstein} present complementary results for a \gls*{HH} dimer, which shows that the changes in the electronic and magnetic correlations functions are zero to numerical precision (note the scale of the $y$-axis in Figs.~\ref{fig:2site_ED_Holstein}b-e). This result demonstrates that the lattice excitations in the \gls*{HH} model are purely phononic in character. All \gls*{ED} results were obtained on two-site half-filled dimers with open boundary conditions while truncating the local phonon Hilbert space to a maximum of $N_\mathrm{p} = 25$ modes per site. 

\newpage
\section{Comparisons of the Hubbard-Holstein and Hubbard-SSH models}
\begin{figure}[h]
    \centering
    \includegraphics[width=0.5\textwidth]{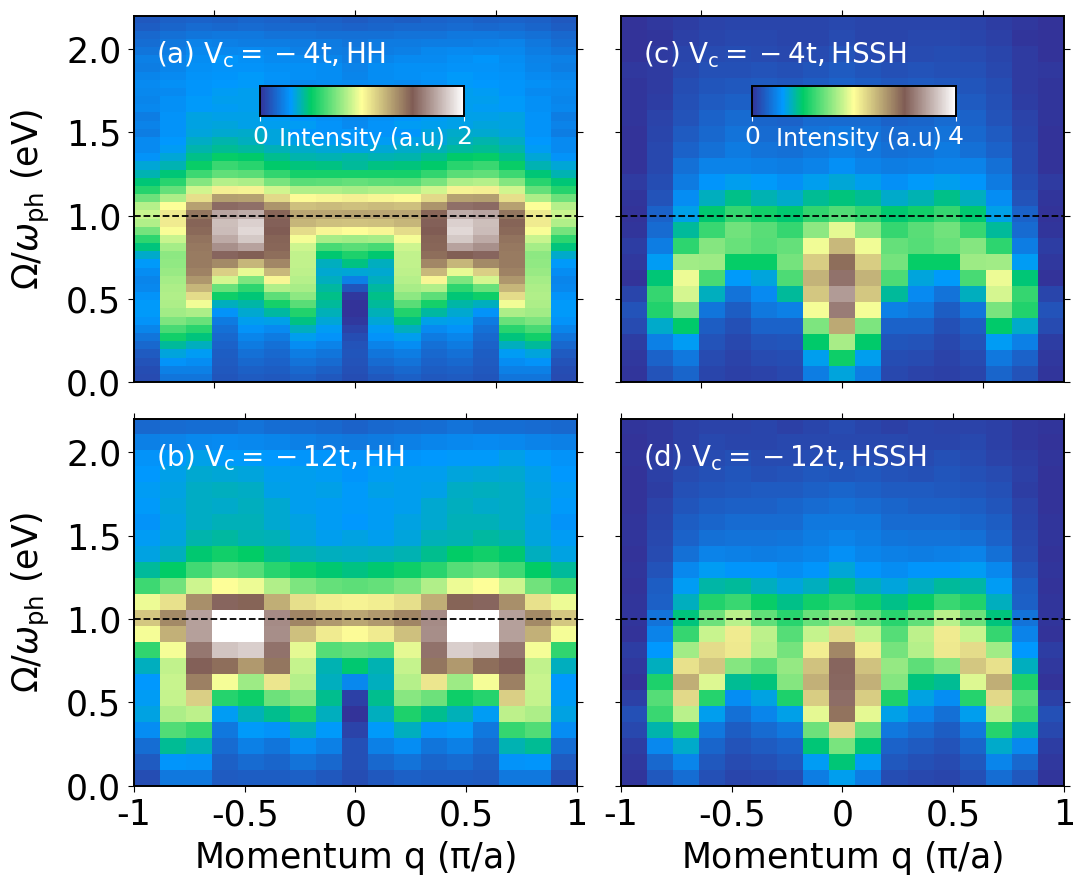}
    \caption{Comparison of the RIXS spectra in the spin conserving channel for the Hubbard Holstein (HH) and Hubbard SSH (HSSH) model as a function of $V_\text{c}$ shown in the panels. All results are obtained with DMRG for a $L=16$ site chain for $U=8t$, $\omega_\text{ph}=t$, $\Gamma/2=t/4$, $g=0.5$ (for HH) and $g=0.4$ (for HSSH). Dashed horizontal lines in the plot show the energy of the bare phonon modes.}
    \label{fig:RIXS_HH_HSSH_DMRG}
\end{figure}

Figure~\ref{fig:RIXS_HH_HSSH_DMRG} compares the spin-conserving \gls*{RIXS} spectra for the \gls*{HH} and \gls*{HSSH} models for $U=8t$, $\omega_\text{ph}=t$, and $\Gamma/2=t/4$. Figs.~\ref{fig:RIXS_HH_HSSH_DMRG}a,b show \gls*{RIXS} spectra for the \gls*{HH} model for $g=0.5$ and varying $V_\text{c}$ as indicated and Fig.~\ref{fig:RIXS_HH_HSSH_DMRG}c,d show the same for \gls*{HSSH} model with $g=0.4$. The spectra for both models have overlapping contributions from magnetic (two-spinon) and lattice excitations. However, for the \gls*{HH} model (Figs.~\ref{fig:RIXS_HH_HSSH_DMRG}a,b), the phonon excitations appears as a dispersionless excitation that tracks the bare phonon energy ($\omega_\text{ph} = t$) across the entire Brillouin zone. Whereas the phonon peaks for the \gls*{HSSH} model (Figs.~\ref{fig:RIXS_HH_HSSH_DMRG}c,d) are redshifted from $\omega_\text{ph}$ due to spin-phonon coupling, as discussed in the main text.

\begin{figure}[h]
    \centering
    \includegraphics[width=0.75\textwidth]{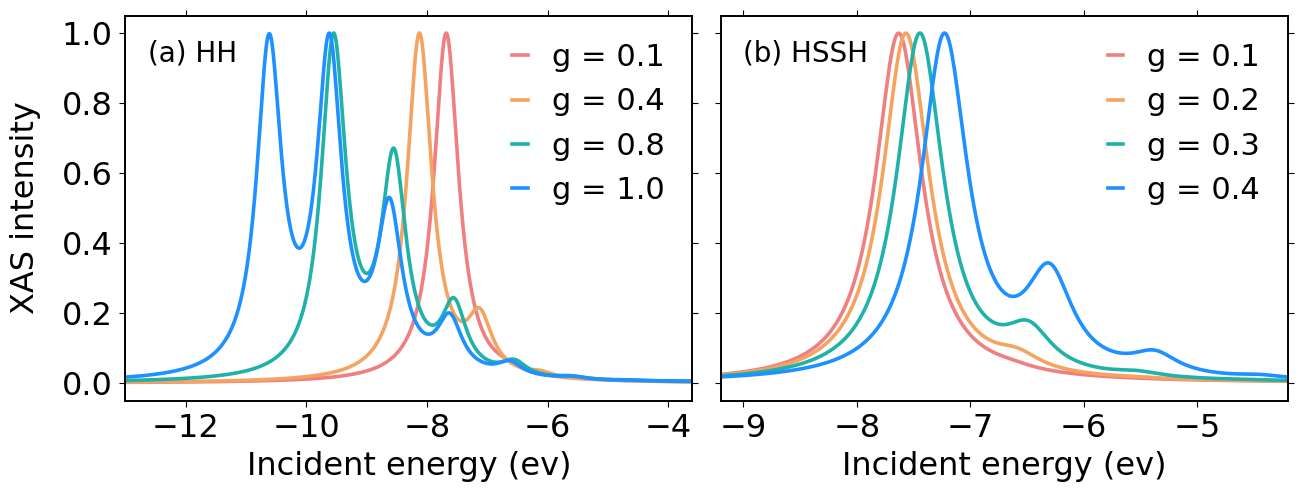}
    \vspace{-0.25cm}
    \caption{The \gls*{XAS} spectra for the (a) Hubbard-Holstein and (b) Hubbard-\gls*{SSH} models. Results were obtained on a two-site dimer using \gls*{ED} for $U=8t$, $\omega_\text{ph}=t$, $V_\text{c}=-8t$, $\Gamma/2=t/4$ and $N_p=25$ phonon modes per site.}
    \label{fig:XAS_HH_HSSH_ED}
\end{figure}

Figure~\ref{fig:XAS_HH_HSSH_ED} compares \gls*{XAS} spectra for the Hubbard-Holstein (HH) and Hubbard-SSH (HSSH) models, calculated with \gls*{ED} on a two-site dimer system, as a function of incident photon energy ($\omega_\text{in}$) for $U=8t$, $\omega_\text{ph}=t$, $V_\text{c}=-8t$, $\Gamma/2=t/4$ and varying $g$ as indicated in the legends. Increasing \gls*{eph} coupling shifts the \gls*{XAS} resonance peaks to lower energies in the \gls*{HH} model (Fig.~\ref{fig:XAS_HH_HSSH_ED}a), whereas an opposite trend is observed for the \gls*{HSSH} model (Fig.~\ref{fig:XAS_HH_HSSH_ED}b). As mentioned in the main text, in the \gls*{HH} model, larger \gls*{eph} interactions increase the binding of the doublon in the intermediate state, which shifts well-screened \gls*{XAS} resonance peaks to lower energies. In contrast, Fig.~\ref{fig:XAS_HH_HSSH_ED}b suggests that \gls*{SSH} interaction is less effective in localizing the excited core electron

\bibliography{references}